# Kerr beam self-cleaning in the multimode fiber anomalous dispersion regime

Y. Leventoux[1], A. Parriaux[3], O. Sidelnikov[6], G. Granger[1], M. Jossent[2], L. Lavoute[2], D. Gaponov[2], M. Fabert[1], A. Tonello[1], K. Krupa[4], A. Desfarges-Berthelemot[1], V. Kermene[1], G. Millot[3], S. Février[1], S. Wabnitz[4,5,6], and V. Couderc[1,*]

[1] *Université de Limoges, XLIM, UMR CNRS 7252, 123 Avenue A. Thomas, 87060 Limoges, France*
[2] *Novae, ZI du Moulin Cheyoux, 87700 Aixe sur Vienne, France*
[3] *Université de Bourgogne Franche-Comté, ICB, UMR CNRS 6303, 9 Av. A. Savary, 21078 Dijon, France*
[4] *Dipartimento di Ingegneria dell'Informazione, Università di Brescia, via Branze 38, 25123 Brescia, Italy*
[5] *Istituto Nazionale di Ottica del Consiglio Nazionale delle Ricerche (INO-CNR), via Branze 45, 25123 Brescia, Italy*
[6] *Novosibirsk State University, 2 Pirogova Street, Novosibirsk, 630090, Russia*

*Corresponding author: vincent.couderc@xlim.fr

**We experimentally demonstrate Kerr beam self-cleaning in the anomalous dispersion regime of graded-index multimode optical fibers. By using 90 ps duration, highly chirped (2 nm bandwidth at -3dB) optical pulses at 1562 nm, we demonstrate a 2 decades reduction, with respect to previous experiments in the normal dispersion regime, of threshold peak power for beam self-cleaning into the fundamental mode of the fiber, accompanied by more than 65% nonlinear increase of intensity correlation into the fundamental mode. Highly efficient self-selection of the LP$_{11}$ mode is also observed. Self-cleaned beams remain spatio-temporally stable for more than a decade of variation of the peak pulse power.**

Multimode optical fibers (MMFs) are extensively investigated for increasing the capacity of future long-distance communications using spatial division multiplexing [1], for nonlinear microscopy and endoscopy applications [2], and for scaling up the energy of fiber laser sources [3]. In applications where MMFs are used for beam delivery, an important problem to be solved is the effective coupling and stable transport of a diffraction limited optical beam from a single mode (SM) laser source into the fundamental mode of the MMF. Random mode coupling limits the MMF length to a few tens of cm, before the beam quality is severely degraded. Although in MMF amplifiers effective SM propagation can be achieved by gain-guiding [4], in passive graded-index (GRIN) MMFs spatial self-cleaning towards the fundamental mode has been only recently demonstrated in the normal group velocity dispersion (GVD) regime, by exploiting the intensity-dependent refractive index of the fiber [5-9].

In this Letter, we experimentally investigate optical pulse propagation in the anomalous GVD regime of GRIN-MMFs. In our experiment, the input beam from a SM fiber laser source leads to the excitation of a few modes of the GRIN MMF, as it occurs for generating a multimode fiber soliton (MMS) when femtosecond pulses are used [10]. Here we use long (90 ps), pre-chirped optical pulses, to avoid dispersive effects to play any role. Still, as we shall see, previously unforeseen spatiotemporal nonlinear attractors, with surprisingly robust stability properties, emerge from nonlinear multimode propagation in the MMF. We found that, depending on the specific input spatial coupling conditions (*i.e.*, incidence angle of the beam), the pulse at the fiber output exhibits a very robust spatial reshaping into either the fundamental LP$_{01}$ or the higher order LP$_{11}$ mode. Thanks to the optimized and nearly SM excitation conditions, the power threshold for Kerr beam self-cleaning is reduced by several orders of magnitude with respect to previous experiments carried out in the normal GVD regime, which used highly multimode input beams.

The reason why a strong pre-chirp of the input optical pulse favors Kerr beam self-cleaning can be understood as follows. In Fig. 1, we present an experimental comparison of the Kerr beam self-cleaning power stability, as a function of the input pulse temporal duration, using various laser sources delivering Fourier transform-limited pulses with different temporal durations at 1064 nm (normal dispersion regime for the MMF). The Gaussian pump beam diameter at the input fiber face is 36 µm (1/e$^2$), corresponding to the excitation of more than 19 transverse modes for all experiments shown in figure 1 (50/125 µm GRIN-MMF). The initial peak power is then sufficiently high to obtain first self-cleaning involving energy transfer into the fundamental mode, followed by self-phase modulation, which is at the origin of the irreversibility of the process [6]. Here we display a series of spectrograms showing, in the vertical coordinate, the x-transverse dimension of the output beam intensity from the GRIN MMF. Whereas the horizontal dimension indicates the corresponding pulse spectrum (wavelength). In each row, the input pulse power increases from left to right. The leftmost column refers to the case of linear pulse propagation: the beam size is wide owing to the speckled output. In the middle column, we show the case corresponding to the power threshold for Kerr beam self-cleaning: as can be seen, the beam size is spatially compressed to the size of the fundamental mode of the MMF, while the spectrum remains nearly as narrow as the initial one. This shows that at the self-cleaning threshold, the self-phase-modulation (SPM) induced spectral broadening is still negligible. Finally, the rightmost column corresponds to the high-power regime, where SPM leads to significant spectral broadening.

In the top row of Fig.1, the input pulse duration was τ=740 ps, which is about 60 times longer than the differential group-delay (DGD) in the MMF. In the bottom row, on the other hand, the input pulse duration was τ=6 ps, that is nearly equal to the fiber DGD. The middle row corresponds to the intermediate case with τ=60 ps or five times the DGD. As can be seen in Fig.1, Kerr beam self-cleaning remains stable (with respect to large variations of the input peak power) in the presence of SPM only in the case of a pulse whose duration is much longer than the DGD, or in the quasi-continuous wave regime, which was recently studied by Krupa et al. [5,6,9]. Therefore, in order to observe Kerr beam self-cleaning, it is necessary that the corresponding power threshold remains lower than the power level where one observes significant SPM-induced spectral broadening. At the same time, the input pulse duration must be much longer than the fiber DGD. Although the results of Fig.1 were obtained with laser sources emitting at 1064 nm, thus in the normal GVD regime of the MMF, the generality of these conclusions does not depend on the sign of chromatic dispersion, but only on the relative magnitude of SPM and DGD with respect to the self-cleaning threshold value (for fiber lengths much shorter than the chromatic dispersion length).

Now, in the anomalous dispersion regime, a MMS in a GRIN fiber is obtained from a balance between SPM and DGD. Simultaneously, each modal component of the MMS remains confined in time thanks to the balance between SPM and GVD. This means that, in order to observe stable Kerr beam self-cleaning in the anomalous GVD regime, it is of advantage to simultaneously use input pulse durations τ which are much longer, and peak powers much lower, than the corresponding values of a MMS. In this way, the effects of both SPM and DGD can be basically neglected. In practice, this can be easily experimentally achieved by means of a dispersive pre-chirping of a short laser pulse, which broadens its time duration while decreasing its peak power. Such low-power, pre-chirped pulse can thus undergo stable self-cleaning in a GRIN MMF. Since its spectral properties are largely unaffected by the spatial self-cleaning process, if so needed, the pulse, in principle, could be subsequently re-compressed to its original duration by a second dispersive element after that self-cleaning has occurred. However, it is also important to check that the pulse temporal envelope does not undergo too much distortion.

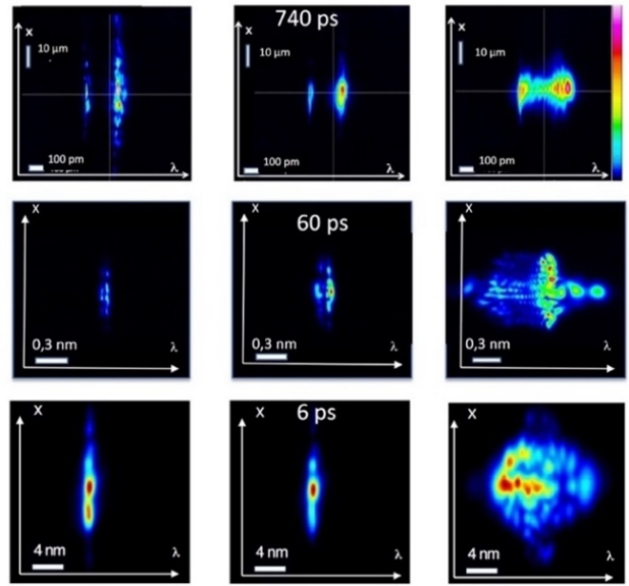

Fig. 1: Spectrograms (intensity vs. transverse coordinate x and wavelength λ) in the near field at the GRIN-MMF output, as a function of the output power (from left to right, 10 W, 1 kW, and 45 kW, respectively) and input pulse width (from top to bottom), for propagation in the normal dispersion regime (50/125 GRIN-MMF, top row: 12 m, middle row: 12 m, last row: 6 m).

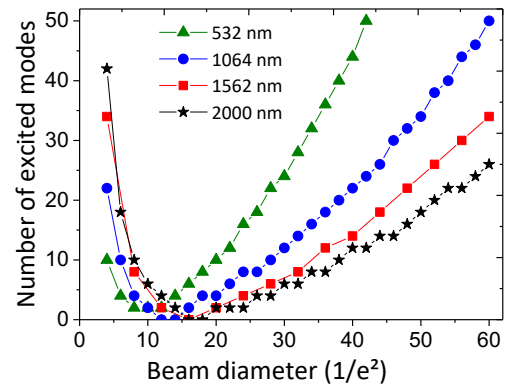

Fig. 2: Number of excited radial modes (with more than 99% of input beam energy) vs. input beam diameter for four different input wavelengths.

The self-cleaning of chirped ultrashort optical pulses is thus analogous to chirped pulse amplification in nonlinear amplifiers [11]: in both cases, the effects of SPM can be reduced by means of pre-chirping the pulses. Nevertheless, in our case, the presence of a nonlinear phase shift between the fundamental mode and high-order modes is required, in order to suppress random mode coupling, and stably maintain the self-cleaned profile.

The experimental set-up to observe chirped pulse Kerr beam self-cleaning in the anomalous GVD regime of a GRIN MMF was based on the dispersive broadening (and chirping) of the full-width at half maximum (FWHM) duration fiber laser pulses from 1 ps to 90 ps. The spectral bandwidth at -3 dB of input pulses was 2 nm. The center of the fiber laser wavelength was 1562 nm, with 1 MHz repetition rate. The beam spot was of Gaussian shape, and ~20 μm in diameter full width at $1/e^2$ in intensity. This leads to optimized free-space coupling efficiency (up to 80%) into the GRIN MMF. The GRIN MMF of 12 m length was loosely coiled on the table forming rings of ~15 cm diameter. The fiber had a circular core of 25 μm radius, with a core-cladding index difference of 0.015 corresponding to a 0.2 numerical aperture. We numerically estimated that these input conditions lead to more than 99% of the pulse energy coupled into the first three radial symmetry modes only (see fig. 2). The diameter and low number of spots in the linear output beam pattern is also the direct signature of energy coupling into a small number of modes (fig. 3). This greatly facilitates the output beam cleaning process, and explains its reduced input power threshold with respect to experiments carried out at 1064 nm with much larger relative input beam diameters (fig.1).

Figure 3 shows the progressive self-cleaning, as the input peak power grows larger, of the output near-field intensity patterns. As can be seen, for average powers above 0.4 mW (corresponding to a peak power of about 4.4 W), the speckled output intensity pattern resulting from the coherent superposition of the multimode field self-cleans into a bell-shaped beam, with negligible multimode field background. We show in Fig. 4 (solid curve), the intensity correlation $C_S$ of the experimental near field pattern of the output beam with the mode $LP_{01}$, numerically obtained with a mode solver:

$$C_S = \frac{\int I_{exp} I_{th} dS}{\sqrt{\int I_{exp}^2 dS \int I_{th}^2 dS}} \qquad (1)$$

where $I_{exp}$ and $I_{th}$ represent the experimental output intensity profile and the numerically calculated mode profile, respectively, and the integration is carried out along the fiber cross section S. The intensity correlation $C_S$ with the fundamental mode keeps monotonically increasing as the input peak power grows larger, until reaching more than 90% at high powers. Additionally, the far and near fields measurement of the output cleaned beam allowed to demonstrate that the beam divergence is only 1.15 times higher than that obtained for a pure Gaussian beam.

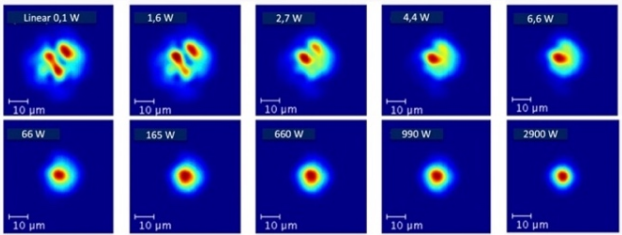

Fig. 3: Near field intensity patterns at the GRIN-MMF output recorded for increasing input peak powers, for appropriate settings of the input coupling, to get the self-selection of a $LP_{01}$ mode.

Figures 3 and 4 show that the process of chirped pulse self-cleaning in our present experiments is much more efficient than in previous experiments carried out in the normal GVD regime, in two respects. First, the threshold for self-cleaning is reduced by about two orders of magnitude (from 0.5 kW [6] down to 4-5 W) for a GRIN fiber length of about 12 m. Second, nearly complete intensity correlation with the fundamental mode is achieved at the fiber output, in contrast with experiments in the normal GVD regime and with a wider input beam, where large residual background was generally observed around the self-cleaned beam [7, 9, 12].

Figures 3-4 show that, in spite of coupling more than 90% of the input beam energy into the fundamental mode of the GRIN MMF, the output beam quality is strongly degraded in 12 m of propagation, owing to linear mode coupling with higher-order modes. In a first step, the Kerr effect acts as a trigger to allow energy transfer between excited modes by means of phase matched four-wave mixing process whereas, in a second step, it prevents from an energy flow out of the fundamental mode. This occurs because the nonlinear nonreciprocity of mode coupling [6]: the nonlinear contribution to the propagation constant of the fundamental mode brings out of phase-matching the random linear mode coupling.

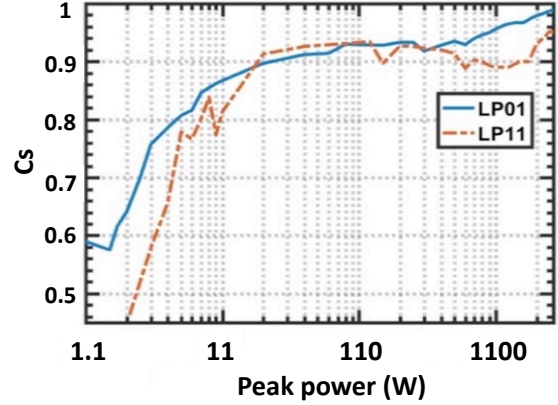

Fig. 4: Intensity correlation $C_S$ upon output peak power for $LP_{01}$ (solid curve) and $LP_{11}$ (dashed curve).

In our experiments, by varying the horizontal tilt of the Gaussian laser beam launched at the GRIN-MMF input face, we could also observe the progressive self-selection of the $LP_{11}$ mode at the fiber output (see Fig.5). The incident beam intercepted the fiber axis with an angle of 2.5° in order to exit the numerical aperture of the fundamental mode. In this case, the coupled power mainly fed odds modes and self-selection was observed above the peak power threshold of about 44 W. Still, the threshold was about two orders of magnitude lower than that for $LP_{11}$ mode self-selection in the normal GVD regime [12]. Even more remarkable is the intensity correlation with the $LP_{11}$ mode, which reaches more than 90% at high powers as illustrated by the dashed curve of Fig. 4.

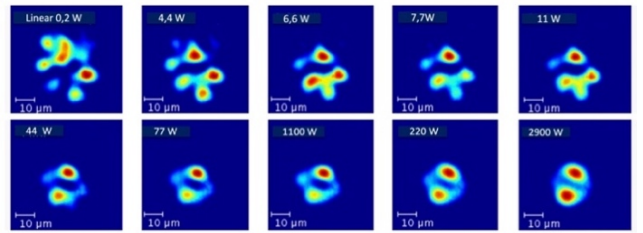

Fig. 5: Near field intensity patterns at the GRIN-MMF output recorded for increasing input peak powers, for appropriate settings of the input coupling, to get the self-selection of a $LP_{11}$ mode.

Besides the stability of the output spatial patterns, the observed nonlinear attractor also exhibits a remarkable stability versus pulse power in both the frequency and the temporal domains. Figure 6 shows the dependence upon of the output peak power of the spectral density of pulses attracted into the mode $LP_{01}$. As can be seen, after a limited spectral compression visible for low input peak power (3.3 W), the output pulse spectrum remains nearly unchanged, in spite of more than two decades of variation of the output average power. SPM is only obtained after the self-cleaning process, and it introduces a significant spectral broadening beyond 100 mW (1100 W peak power). For a maximum of pump power, Raman scattering comes into play and it introduces a spectral asymmetry towards the highest wavelengths. Then, additional cleaning brought by the Raman gain can additionally improve the output beam profile. The corresponding output temporal intensity profiles measured by a 70 GHz oscilloscope (5 ps of time resolution) and a fast photodiode placed in the center of the output beam pattern show significant reshaping. Such evolution is the signature of energy exchange among modes which do not have the same transverse energy repartition. Then, the energy relocation can significantly affect the pulse envelope profile, and produce a shift of the pulse peak as high as 90 ps, when increasing the input power up to 265 mW (2915 W peak power) [9]. At the maximum pump power, the output pulse duration nearly doubles with respect to the input value at low powers.

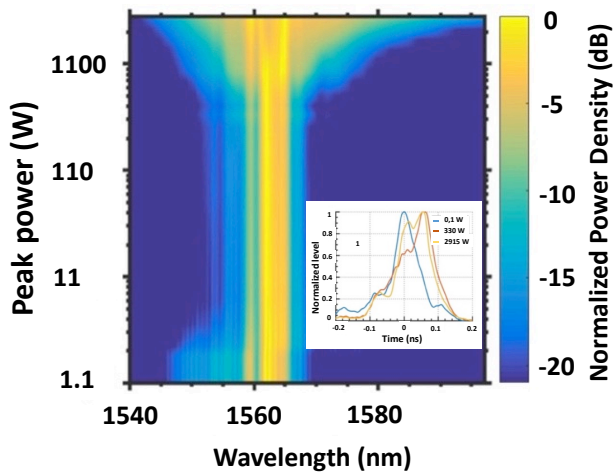

Fig.6: Experimental power spectral density (normalized to its local peak value) of pulses at the output of the GRIN MMF, as a continuous function of the output peak power. Inset: normalized pulse intensity for three selected values of the output peak power.

To conclude, we have experimentally shown that Kerr beam self-cleaning in GRIN-MMFs can occur with extreme efficiency in the anomalous dispersion regime. By controlling the input coupling conditions, nearly full reshaping into either the fundamental $LP_{01}$ or higher-order $LP_{11}$ mode of the MMF has been observed. The threshold for beam self-cleaning is reduced by about two orders of magnitude with respect to previous experiments in the normal dispersion regime. This significant improvement is due to the input beam conditions, which have been properly controlled to excite a low number of transverse modes. The dispersion regime does not appear to be a key parameter, as long as the initial peak power remains sufficiently low, so to avoid additional nonlinear effects (such as modulation instability or soliton generation). Self-cleaning, either obtained on the $LP_{01}$ or in the $LP_{11}$ mode, can be driven by the initial conditions governing the excitation of either even ($LP_{01}$) or odd modes ($LP_{11}$). The self-cleaned beams remain nearly unchanged in the frequency domain, for over two decades of variation of the output power, whereas the pulse envelope undergoes significant evolution under energy exchange between modes. These results are relevant for applications of self-cleaning in multimode fibers in different fields, such as three-photon imaging in medical microscopy and endoscopy, for the development of high-power mode-locked multimode fiber lasers and amplifier operating in the near-infrared, and for space division multiplexed transmission with GRIN fibers.

**Funding.** The European Research Council (ERC) under the European Union's Horizon 2020 research and innovation programme (No. 740355). K.K. has received funding from the European Union's Horizon 2020 research and innovation programme under the Marie-Skłodowska-Curie (No. 713694). Agence Nationale de la Recherche (ANR) (Broadband Infrared Supercontinuum BISCOT ANR-16-CE08-0031); Région Nouvelle Aquitaine (FLOWA). Direction Générale de l'Armement (DGA), french program "Investissements d'Avenir", project ISITE-BFC (contract ANR-15-IDEX-0003). O.S. received funding from the Russian Ministry of Science and Education (Grant 14.Y26.31.0017).

**Acknowledgment**. We acknowledge helpful discussions with Alain Barthélémy.